\documentclass{emulateapj}
\usepackage{epsfig}
\usepackage{amssymb}
\usepackage{graphicx}

\slugcomment{To appear in ApJ Letters}

\shorttitle{\emph{Herschel} Observations of the T Cha  Transition Disk}
\shortauthors{Cieza, L. et al.}

\begin{document}

\title{\emph{Herschel}  Observations of the T Cha Transition Disk: constraining  the outer disk properties}

\author{Lucas A. Cieza\altaffilmark{1},
Johan Olofsson\altaffilmark{2},
    Paul M. Harvey\altaffilmark{3},
    Christophe Pinte\altaffilmark{4},
        Bruno Mer\'in\altaffilmark{5},
    Jean-Charles Augereau\altaffilmark{4}, 
    Neal J. Evans II\altaffilmark{3},
    Joan Najita\altaffilmark{6},
    Thomas Henning\altaffilmark{2}, and
    Francois M\'enard\altaffilmark{4}}      

\altaffiltext{1}{Institute for Astronomy, University of Hawaii at Manoa,  Honolulu, HI 96822. \emph{Sagan} Fellow, lcieza@ifa,hawaii.edu}
\altaffiltext{2}{Max Planck Institut f\"ur Astronomie, K\"onigstuhl 17, 69117, Heidelberg, Germany}
\altaffiltext{3}{Department of Astronomy, University of Texas at Austin,  Austin, TX 78712}
\altaffiltext{4}{UJF-Grenoble 1 / CNRS-INSU, Institut de Plan\'etologie et d'Astrophysique de Grenoble (IPAG) UMR 5274, Grenoble, F-38041, France}
\altaffiltext{5}{Herschel Science Centre, European Space Agency (ESAC), P.O. Box, 78, 28691 Villanueva de la Ca\~{n}ada, Madrid, Spain}
\altaffiltext{6}{National Optical Astronomy Observatory, 950 N. Cherry Avenue, Tucson, AZ 86719, USA}

\begin{abstract}

\noindent
T~Cha is a nearby  
(d $\sim$ 100 pc) 
transition disk known to have an optically thin gap separating  optically thick inner 
and outer disk components.  Hu\'elamo et al. (2011) recently reported 
the presence of  a low-mass object candidate  within  the gap of  the T~Cha disk, giving credence to the  suspected planetary origin of this gap.
Here we present the  \emph{Herschel}  photometry  (70, 160, 250, 350, and 500 $\mu$m) of T~Cha from the  ``Dust, Ice, and Gas in Time" (DIGIT) Key Program,  which  bridges the wavelength range between existing  \emph{Spitzer} and  millimeter data
and provide important constraints on the outer disk properties of this extraordinary system. 
We model the entire optical to millimeter wavelength spectral energy distribution (SED) of  T~Cha (19 data points between 0.36 and 3300 $\mu$m without any major gaps in wavelength coverage).
T Cha shows a steep spectral slope in the far-IR,  
which we find clearly favors models with outer disks containing little or no dust beyond $\sim$40 AU. 
The full SED can be modeled equally well with either an outer disk that is very compact (only a few AU wide) or a much larger one 
that has a very steep surface density profile. That is,  T~Cha's outer disk seems to be either very small or very tenuous. 
Both scenarios suggest a highly unusual outer disk and have important but different implications for the nature of  T~Cha. 
Spatially resolved  images are needed to distinguish between the two scenarios.   
\end{abstract}


\section{Introduction}

Transition disks were first identified by the \emph{Infrared Astronomical Satellite}  as objects with little or no excess
emission at $\lambda$ $<$ 10 $\mu$m and a significant excess at $\lambda$ 
$\ge$ 10 $\mu$m  (Strom et al. 1989). 
This broad definition encompasses a wide range of SED morphologies (Najita et al. 2007).  
T Cha belongs to a subclass of transition disks known as cold disks, which are characterized by a \emph{steep} rise 
in the SED between  $\sim$10 and $\sim$ 30 $\mu$m (Brown et al. 2007; Mer\'in et al. 2010).  It is now well 
established, from both SED modeling 
and direct (sub)millimeter imaging,  that the  peculiar SED shapes of  cold disks are due to the presence of large inner holes 
or gaps in their dusty disks.
Inner holes can be produced by a  number of processes, including grain growth, photoevaporation, and dynamical interactions with  
(sub)stellar or planetary mass companions (see Williams $\&$ Cieza, 2011 for a recent review).   
However, objects such as T~Cha,  LkCa 15, and UX Tau are believed to have an optically thick  inner disk 
within a much larger inner hole (Olofsson et al. 2011, Espaillat et al. 2010).
Optically thin gaps separating  optically thick inner  and outer disk components are difficult to explain by anything other 
than the dynamical interaction of substellar or planetary mass objects  embedded within the disk. 
The discovery of what seems to be either a young  planet or a brown dwarf surrounded by a significant amount of dust
within the gap of  the T Cha disk (Hu\'elamo et al. 2011) supports the suspected planetary origin of the gap.  
Detailed studies of the properties of the  T Cha  disk thus become highly desirable to further 
our understanding of protoplanetary disk evolution and planet formation.  
Using  \emph{Spitzer} photometry and  spectroscopy, Brown et al. (2007) first modeled T~Cha 
 with an inner disk extending from 0.08 to 0.2 AU and an outer disk  extending from 15 to 300 AU.
More recently, Olofsson et al. (2011) simultaneously modeled the SED of T Cha and near-IR interferometric observations
obtained with the AMBER instrument at the Very Large Telescope Interferometer (VLTI; Scholler
2007). 
They spatially resolved the narrow inner dusty disk and constrained its inner edge to be $\sim$\,0.1\,AU from the star.

Here we present the \emph{Herschel}  photometry (at 70, 160, 250, 350, and 500 $\mu$m) of T Cha and model its
entire optical to millimeter wavelength SED.  We focus our analysis on the properties of the outer disk (i.e., beyond the gap),
which is the region that is both the best probed by \emph{Herschel} data and the least constrained by previous studies.  

\section{Observations}

\subsection{\emph{Herschel} observations and data reduction}

We obtained far-IR and submillimeter  wavelength photometry for  T Cha as part of the DIGIT Open Time \emph{Herschel}  
Key  Program
using both the Photodetector Array Camera and Spectrometer (PACS, 70 and 160 $\mu$m, Poglitsch et al.  2010) and the Spectral and Photometric Imaging Receiver
(SPIRE, 250, 350 and 500 $\mu$m, Griffin et al. 2010).  The PACS observations (OBS ID = 1342209063)  were obtained  on November 5$^{th}$, 2010, using the
Mini Scan Map  mode, a repetition factor of 3, and a scan angle of 70 degrees. 
Similarly,  the SPIRE observations (OBS ID =  1342203636) were taken on August 24$^{th}$, 2010,  using the  Small Scan Map mode,
and a repetition factor of 3. 
Both the PACS and SPIRE data were processed using HIPE (Herschel Interactive Processing 
Environment; Ott et al. 2010) version 7.1.  The PACS beam sizes at 70, and 160 $\mu$m are 5.5$''$ and 11$''$,
and the data were resampled to 1$''$ and 2$''$ pixel images, respectively. SPIRE beam sizes at 250, 350 and 
500 $\mu$m are  18.1, 25.2 and 36.6$''$, respectively, and  the images were resampled to 6$''$, 10$''$, and 14$''$ pixel images,
respectively. 

T Cha was detected with good signal to noise ( $\gtrsim$ 15-50) at all PACS and SPIRE wavelengths.
PACS and SPIRE fluxes were derived with a psf-fitting routine essentially identical to
the c2dphot software used for the {\it c2d Spitzer} Legacy Program (Evans et al. 2007).  
We used the empirical psf's posted by the NASA Herschel Science Center\footnote{https://nhscsci.ipac.caltech.edu/sc/index.php/}
and found that they gave a good match to the actual images of T Cha.  In addition, we checked that aperture photometry gave similar
results and assigned the final uncertainties based on the 10$\%$ and 15$\%$ error 
estimated for  PACS and SPIRE absolute photometry, respectively. 
We also  added an additional 5$\%$  uncertainty to reflect incomplete knowledge
of the psf and the agreement between the psf-fitting and the aperture photometry.
The resulting \emph{Herschel} photometry for T~Cha is listed in Table 1. 

\subsection{Photometry data from the literature}

\begin{figure}
\includegraphics[width=16cm, trim = 0mm 0mm 0mm 150mm, clip]{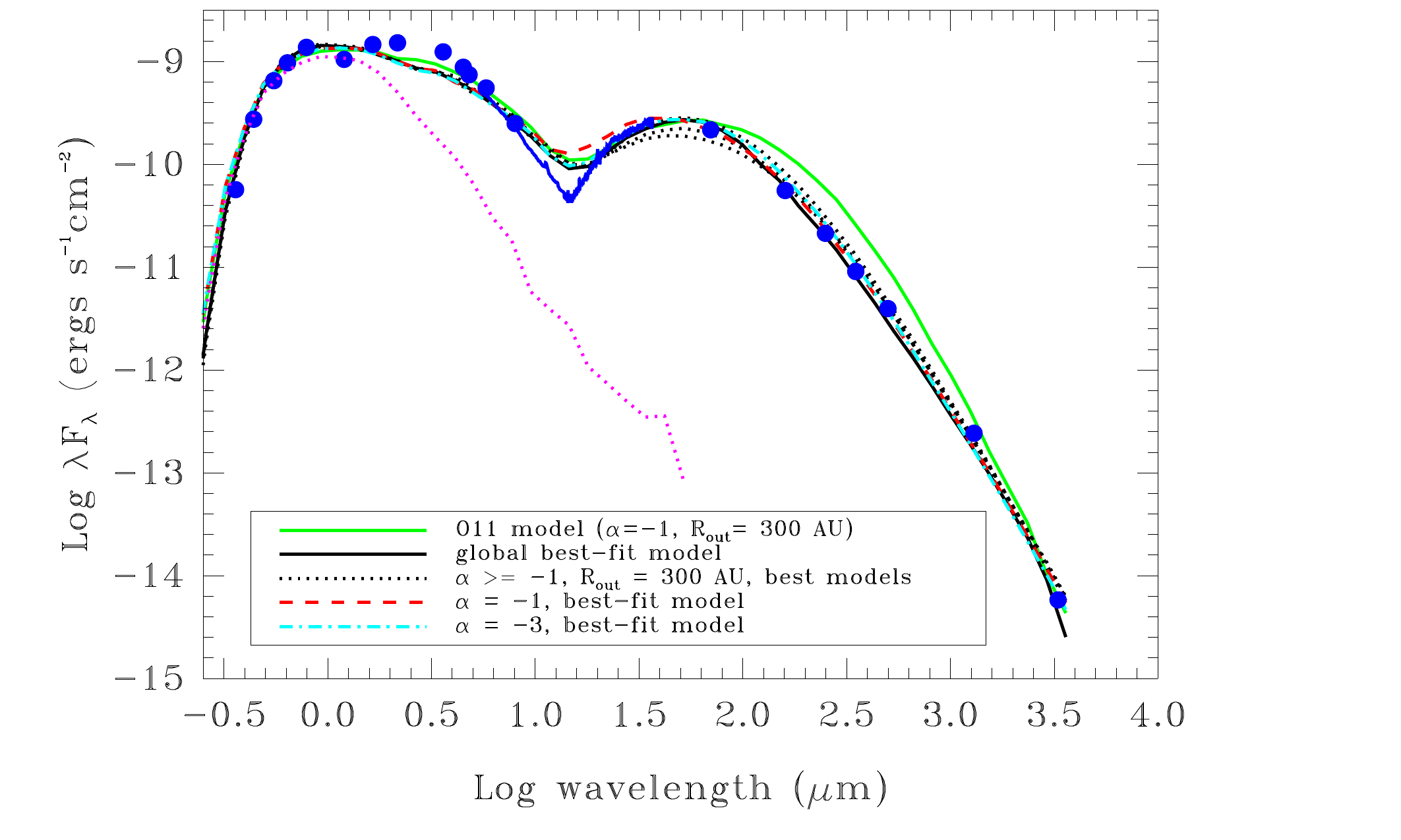}
\includegraphics[width=15cm,  trim = 10mm 00mm 0mm 10mm, clip]{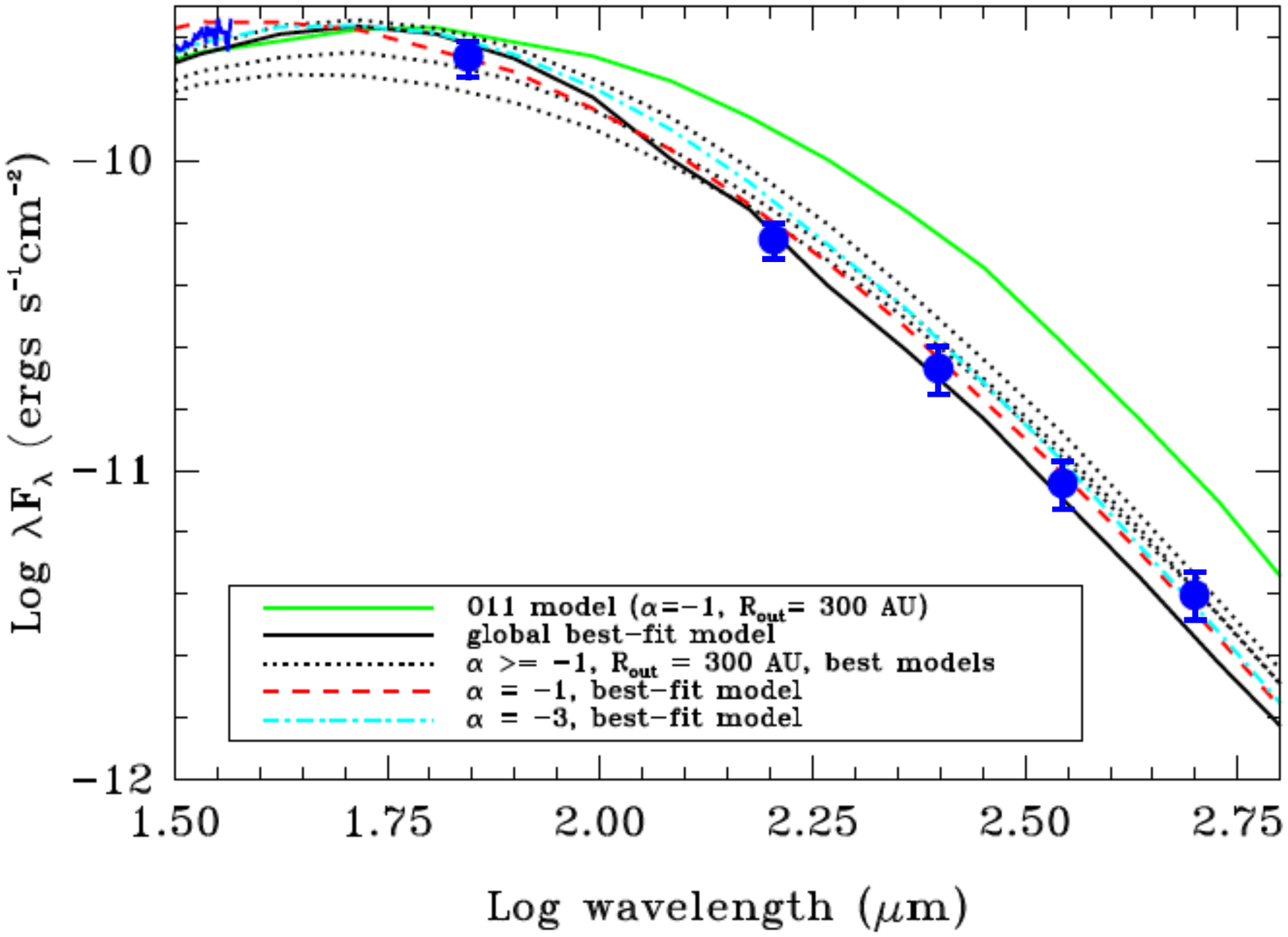}
\caption{ \small 
\textbf{Top Panel:} The full Spectral Energy Distribution of T Cha and a variety of models as labeled.
The O11 model fits all previously available data but greatly overestimates the new 160, 250, 350, and 500 $\mu$m
fluxes, illustrating the power of \emph{Herschel} to completely rule out certain models. 
\textbf{Lower Panel:} The SED region sampled by \emph{Herschel}. 
While the global best-fit model has a
far-IR slope slightly steeper than observed, 
all disk models with a  shallow surface density profile 
and a large outer radius, including the O11 model, 
result in far-IR colors that are much too red.
The surface density profile of the outer disk and its inner and outer radii are degenerate parameters.
Equally good fits  can be obtained by  either adopting a small outer radius (R$_{out}$ = 30 AU, red dashed line) 
or a very steep surface density profile ($\alpha$ = $-$3, light blue dot-dashed line). 
}
\label{fig:sed-mix}
\end{figure}

In order to construct the full SED of T Cha, we collected optical, near-IR, mid-IR,  and mm wavelength
photometry from the literature, which are also listed in Table 1.  
The full optical to mm wavelength  SED for  T ~Cha  is shown in Figure~\ref{fig:sed-mix}.   
There are now a total of 19  photometry points  without major gaps in wavelength  coverage between 0.36  and 3300 $\mu$m. 
In addition to the above photometry,  the T Cha SED is well sampled by a \emph{Spitzer}-IRS spectrum, covering the 5.2 to 38 $\mu$m 
region (Brown et al. 2007). The sharp dip in the SED around 15 $\mu$m indicates the presence of a wide gap in the disk. 

\section{Disk model}\label{disk-model}

The starting point of our study is the model presented by  Olofsson et al. (2011, O11 hereafter).
Following O11, we use the Monte Carlo radiative transfer code MCFOST (Pinte et al.
2006; 2009).
We parameterize the structure of the T Cha disk with the following parameters:
 the inner and outer radii  (R$_{in}$ and  R$_{out}$, respectively), the index
 $\alpha$ for the surface density profile
  ($\Sigma$
(r) = $\Sigma$ (r/r$_0$)$^{\alpha}$ ) and a disk scale height $H(r)$
 assuming a vertical gaussian distribution (exp[$-$z$^2$/2H(r)$^2$]),  appropriate for
 a disk in hydrostatic equilibrium. The disk's flaring
is described by a power-law determining the scale height as a
function of the radius ($H(r)$ = H$_0$(r/r$_0$)$^{\beta}$). 
The dust content is
described by a differential power-law for the grain size distribution
(d$n$(a) $\propto$ a $^{p}$d$a$), between the minimum (a$_{min}$) and maximum
(a$_{max}$) grain sizes. 
The total mass of dust in the disk with grain sizes between  a$_{min}$ and 
a$_{max}$ is given by the M$_{dust}$  parameter. 
Each of the above parameters can be defined independently for the  inner and 
outer disk components, which are separated by a wide gap. 

The main  goal of this Letter is to investigate the constraints the \emph{Herschel}
data provide on the outer disk properties of T Cha. 
As the \emph{Herschel} data give no new information on the inner regions, we adopt the 
inner disk parameters from O11, which are listed in Table 2, as fixed values.  
The stellar parameters for T Cha are also taken from O11 and 
are listed in Table 2 as well.  We also fix the minimum  grain
size for the outer disk and its inclination.  We are thus left with 7 free model parameters,
all of which are related to the outer disk: R$_{in}$, R$_{out}$, $\alpha$, $\beta$, a$_{max}$,
M$_{dust}$, and H$_{50}$ (the scale height at 50 AU). 
Motivated by the lack of 10 $\mu$m silicate emission in the \emph{Spitzer}-IRS spectra  of T Cha, 
O11 adopted for the inner disk a mixture of astro silicates  (Draine $\&$ Lee 1984)  and amorphous carbon 
(Zubko  al. 1996) with a mass ratio of 4:1 between silicate and carbon.
For simplicity, we adopt the same silicate and carbon mix for the dust composition of the outer disk,
but remind  the reader that disk parameters derived from SED fitting are strongly dependent on the choice 
of dust properties. 
Although discussing the dependence of each parameter on dust 
composition is beyond the scope of this Letter, we have
verified that our main conclusions remain unchanged if pure
astro silicate grains are adopted for the outer disk.

\subsection{Model grid and best-fit models} 

In order to constrain the parameter space consistent with the observed T Cha SED,  we ran a grid of
models varying the 7 free parameters in our model setup.  The values sampled by the grid are
listed in Table 3.  
The parameters $\beta$ and H$_{50}$ are sampled linearly, while
R$_{in}$, and R$_{out}$, $\alpha$,  a$_{max}$, and M$_{dust}$ are sampled logarithmically. 
The number of values per parameter range from 3 to 10 and the total number of models
amount to 126000. 
The parameters of the best-fit model (i.e., with the minimum $\chi$$^2$)
are also listed in Table 3.
The global  best-fit model  so derived for the outer disk of T Cha is a narrow ring with an inner radius of $\sim$18 AU 
and a width of only $\sim$2 AU. This narrow ring remains optically thick even at (sub)millimeter
wavelengths, which explains the very large mass that is allowed by the model. 
Its SED is shown as a solid black line in Fig. 1.

Since SED modeling is known to be highly degenerate, the best-fit model is unlikely to be a unique solution. 
We therefore adopt the Bayesian method (Press et al. 1992; Pinte et al. 2008) 
to calculate the probability of  different values for each parameter given the available data. 
We assume that we have no prior knowledge on the model parameter, which implies that the relative 
probability of a given model is proportional to exp$[-\chi^2/2]$. 
All probabilities are normalized so the sum of all the probabilities of the models in
the grid is equal to 1.  
The top two panels in Fig. 2 show the probability that  each parameter takes a given value from the grid, 
calculated by summing over the probabilities of all the other parameters. 
Clearly, the  M$_{dust}$, a$_{max}$, and  R$_{in}$ parameters are much better constrained than the others. 
The lower panels in Fig. 2 show the joint probabilities of the R$_{in}$, R$_{out}$, and $\alpha$
parameters. Models with R$_{in}$ $\sim$ 18 AU and slightly larger R$_{out}$ values 
are clearly favored to best reproduce our observations.
There are a large number of narrow ring models that can reproduce the observed T~Cha SED 
relatively well.  The best-fit model is an instance of a large family of such models. 
This can be understood considering that the SED of a narrow ring is little sensitive 
to parameters such as $\alpha$, $\beta$, or even M$_{dust}$ and a$_{max}$.
Wider rings are also allowed, but they require fine-tuning of other parameters (see for example the red dashed line in in Fig. 1 
and the $\alpha = -1$ model in Table 3) and  thus appear as less likely models in the joint probability plots.  
Large disks (R$_{out}$ $\gtrsim$ 150 AU) with shallow surface density profiles ($\alpha \gtrsim$ -1)  result in a poor fit to the SED. This can be  
seen in Fig. 2 in  the dark region  in the upper right corner of the $\alpha$ vs R$_{out}$ plot 
(lower row) and in  the red histogram in the $\alpha$ probability distribution (middle row). 
We find that  all such models  result in far-IR \emph{Herschel} colors  that are much redder than observed (see bottom panel in Fig. 2).  
In other words, all disk models with a  shallow surface density profile ($\alpha \gtrsim -1$) and a
large outer radius (e.g.,  $\gtrsim$ 150-300 AU) either under-predict the observed 30 and 70 $\mu$m fluxes and/or 
over-predict  the 250 and 350 $\mu$m fluxes.
This  is not surprising considering  that a narrow ring around T Cha (like the best-fit model)
would be ``missing" most of the cool material emitting at 250 and 350 $\mu$m 
expected  for a large disk  with a relatively shallow surface density profile.  
However, a narrow ring is not the only way to reproduce the far-IR \emph{Herschel} colors. 
We also find that the outer disk can be arbitrarily  large (e.g., R$_{out}$ = 300 AU)
as long as the surface density profile is sufficiently steep (e.g.,  $\alpha = -3$) 
to quickly reduce the amount of dust at large radii. 
See for example the alternative $\alpha = -3$ model in Table 3, corresponding to the light 
blue dot-dashed line  in Fig. 1. 
This alternative model falls outside of  our parameter grid, but illustrates well the strong 
degeneracy between the outer radius and the surface density profile of the disk.
We note that while the $ \alpha = -1$ and $\alpha = -3$ models have very  different  
R$_{out}$ values, their  M$_{dust}$, R$_{in}$,  a$_{max}$, $\beta$, and  
H$_{50}$  parameters are much more similar. 

Independently of the probability space they occupy in Fig. 2, 
the 3 models discussed above  and listed in Table 3 are fully consistent with all the available data and, 
\emph{individually}, can be considered to be equally good descriptions of the T Cha disk. 
One feature that our models cannot reproduce well is the pronounced V-shape of the \emph{Spitzer}-IRS spectra around 15
$\mu$m.
The  models also underestimate the 2.2 and 3.6 $\mu$m excesses from the inner disk, which parameters 
are heavily driven by near-IR interferometry observations. 
Near-IR variability makes it difficult to simultaneously fit multi-epoch observations in this spectral region
(see O11 for a detailed description of the  inner disk modeling).
Because the inner disk is optically thick, the exact geometry of the inner disk might affect some of the
outer disk parameters like H$_{50}$ and R$_{in}$ (Mulders et al. 2010), but is unlikely 
to play any role on the values of $\alpha$ and R$_{out}$, the key parameters in this study. 

\section{Discussion}

The results from the previous section show that T Cha SED can be modeled with
multiple combinations of parameter values. The most likely models can be divided into 
two broad categories: narrow rings and wider disks with steep surface density profiles.
Both families of models suggest a highly unusual outer disk and have important 
but different implications for the nature of  T~Cha. 
A narrow ring would strongly suggest that the  outer disk has been truncated by the  dynamical interaction 
of a companion, as is known to be the case in the HD 98800B and Hen 3-600A multiple systems, which have 
outer radii of  only 10 and 20 AU, respectively (Andrews et al. 2010). 
Very deep imaging of  T Cha (Chauvin et al. 2010; Vicente et al. 2011) rule out any \emph{stellar}
companion that could be responsible for the truncation of
the outer disk, unless its \emph{projected} separation is currently less than 0.1$''$ (10 AU).
Furthermore, recent aperture masking observations by Hu\'elamo et al. (2011) only
report a very low-mass companion candidate  (either a young  planet or a brown dwarf) 
with a \emph{projected} separation of 6.2$\pm$0.7 AU. 
Since this object is the most likely  cause of the inner cavity (R$_{in}$ $\sim$ 10-20 AU based on
the SED modeling), its \emph{physical}  separation  must be  $<$ R$_{in}$.  
Any object truncating the outer disk thus must be substellar or planetary in nature. 
We speculate that the recently reported companion to T Cha could potentially also provide 
the explanation  for the models with $\alpha \lesssim -2$.  
These unusual $\alpha$ values would imply  that the disk  material is highly concentrated toward the inner 
edge of the outer disk (i.e., that material is accumulating at the outer edge of the gap, at $\sim$15 AU). 
As a reference,  $\alpha  =  -1.5$ in the  canonical  Minimum Mass Solar Nebula (Weidenschilling, 1977).
T~Cha's outer disk is relatively massive but  shows little or no accretion
onto the star (Alcala et al. 1993,  Schisano et al.  2009).
The transport of angular  momentum  and material across  the 
outermost part of the disk should not be affected by the presence of the
low-mass object,  
but when circumstellar material reaches the outer edge of the gap, it can be accreted by
the low-mass object, channeled toward the inner disk, or accumulated at 
the gap's edge  (Lubow \& D'Angelo,  2006).   
The dynamical interaction between a gas-rich disk and a planetary mass object 
 is a complex  hydrodynamic problem  that  depends on the mass 
 and location of the  planet as well as on the properties of the disk  (viscosity, disk  scale height,  etc), 
 but  an $ \alpha$ value of $\lesssim -2$ would suggest that  it could have an important 
 effect on the surface density  of the outer disk.  
Recent 3-D hydrodynamic models of planet-disk interactions 
show  the accumulation of   mm-sized grains in the pressure maxima 
of the gas caused by the spiral density  waves triggered by the planet
(Fouchet et  al. 2010). This implies that the  density enhancement at the edge 
of the gap of the large grains traced by  \emph{Herschel} could be even
larger than it is for the gas.  
However, given the degeneracy between $\alpha$
and R$_{out}$, the dynamical effect  of the companion on the disk cannot be tested without 
spatially resolved observations.
Our models show that, given the SED data alone
 and without knowing R$_{out}$, 
 the value of $\alpha$  for the dust disk could be \emph{anywhere} between 
 $-0.75$ and $-3$. 

\section{Conclusions}

New \emph{Herschel} observations of T Cha reveal a steep spectral slope between 70 and 350~$\mu$m 
(i.e., very blue  far-IR \emph{Herschel} colors compared to a ``normal disk"). 
The steeply decreasing far-IR SED is \emph{not} due to a lack of large grains. 
In fact, the (sub)millimeter  wavelength colors are rather red, and
grains larger than $\sim$1~mm are needed to simultaneously fit 
the \emph{Herschel} and 1.3 and 3.3~mm fluxes. 
Instead, we find basically two ways to reproduce the observed far-IR \emph{Herschel} colors: 
making the outer disk very small (by reducing the value of R$_{out}$) 
or very tenuous (by reducing the value of the surface density profile,  $\alpha$). 
Both types of models contain little or no dust beyond $\sim$40 AU,  but each family 
has  a different astrophysical implication:  either T~Cha has a small  outwardly 
truncated disk  or material is accumulating at the inner edge  of the gap.  
Resolved images are mandatory to break the degeneracy between 
$\alpha$, and the  inner and outer disk radii. 
The Atacama Large Millimeter/submillimeter Array (ALMA) will have both the resolution
and sensitivity needed to directly measure the width of the disk as well
as its surface density profile  and distinguish between the two interpretations. 
The former result would encourage even deeper searches for a wide-separation
low-mass companion (most likely a Jovian planet).
The latter scenario would have  important implications for
the dynamical interaction between the T Cha disk and the recently reported companion 
as well as on any subsequent planet formation in its highly peculiar outer disk.

\acknowledgments
Support for this work, part of the DIGIT \emph{Herschel} Open Time Key
Program, was provided by NASA through an award issued by JPL/Caltech. 
LAC acknowledges support from  NASA through
the \emph{Sagan} Fellowship Program.
JO acknowledges a grant from the Alexander von Humboldt Foundation.
CP, JCA, and FM  acknowledge funding from the 
European Commission's 7$^\mathrm{th}$ Framework Program 
and the Agence Nationale pour la Recherche of France
(projects  FP7: PERG06-GA-2009-256513, ANR-07-BLAN-0221, ANR-2010-JCJC-0504-01)
and Programme National de Physique Stellaire (PNPS) of CNRS/INSU, France.

\newpage

\begin{deluxetable}{crrlr}
\footnotesize
\tablecaption{T Cha Photometry data}
\tablehead{\colhead{Wavelength}&\colhead{Flux}&\colhead{Error}&\colhead{Telescope}&\colhead{Reference$^a$}\\
                      \colhead{($\mu$m)}&\colhead{ (mJy) }&\colhead{(mJy)}&\colhead{}&\colhead{}}
\startdata
  0.36        &     6.80e+00 &   10$\%$  &    ESO 50 cm   &   1 \\
  0.44        &     4.03e+01  &  10$\%$  &   ESO 50 cm &  1 \\
  0.55         &    1.12e+02  &   10$\%$ &    ESO 50 cm &  1 \\
  0.64         &    2.01e+02  &    10$\%$ &   ESO 50  cm  & 1 \\
  0.79        &     3.61e+02   &   10$\%$ &   ESO 50 cm & 1 \\
  1.2          &  4.20e+02  &     10$\%$  & 2MASS  &   2 \\
  1.6          &  8.05e+02   &   10$\%$   &   2MASS & 2 \\
  2.2          &   1.10e+03   &   10$\%$      &  2MASS & 2 \\
  3.6         &  1.49e+03  &   10$\%$   & \emph{Spitzer} & 3 \\
  4.5         &   1.32e+03  &   10$\%$  & \emph{Spitzer} & 3 \\
  5.8         &   1.07e+e3  &   10$\%$  & \emph{Spitzer} & 3 \\
  8.0         &   6.66e+02  &   10$\%$  & \emph{Spitzer} & 3 \\
  70  &   5.06e+03  &  15$\%$  &  \emph{Herschel}   &  4\\
160  &   2.97e+03  &  15$\%$  &  \emph{Herschel}  &  4 \\
250  &   1.78e+03  &   20$\%$   &  \emph{Herschel}  & 4 \\
350  &    1.06e+03  &  20$\%$   &  \emph{Herschel}  &  4\\
500  &    6.60e+02  &  20$\%$  &  \emph{Herschel}  &  4\\
1300 &   1.05e+02   &  15$\%$  &          SEST             &  5 \\ 
3300 &    6.40e+00   &  15$\%$   &        ATCA             & 5 \\
\enddata
\tablecomments{$^a$References:  (1) Alcal\'a et al. 1997;  (2) Skrutskie et al.  (2006);  (3)  Brown et al. (2007); (4) this work;   (5) Lommen et al. (2007)}
\end{deluxetable}

\begin{deluxetable}{lr}
\footnotesize
\tablecaption{Fixed Parameters from O11}
\tablehead{\colhead{Parameter}&\colhead{Adopted value}} 
\startdata
Stellar T$_{eff}$         [K]                       & 5400     \\
Stellar radius                    [R$_{\odot}$]  &     1.3 \\
Stellar mass   [M$_{\odot}$]   &   1.5 \\ 
Distance        [pc]                           &   100 \\
Inclination     [deg]                         &  60 \\
Grain size distribution slope, p   &  $-$3.5     \\
Inner disk R$_{in}$     [AU]     & 0.13  \\
Inner disk R$_{out}$  [AU]        &0.17    \\
Inner disk M$_{dust}$ [M$_{\odot}$] &  10$^{-9}$ \\
Inner disk a$_{min}$  [$\mu$m ] &  0.1     \\
Inner disk a$_{max}$ [$\mu$m] & 10        \\
Outer disk a$_{min}$  [$\mu$m ] &  0.01     \\
\enddata
\end{deluxetable}

\begin{deluxetable}{lrrrr}
\tablecaption{Model Grid and Fitted Outer Disk Parameters}
\tablehead{\colhead{Parameter} &   \colhead{values sampled by grid}   &  \colhead{best-fit  model} &\colhead{$\alpha$ =  $-$1 model$^a$}&\colhead{$\alpha$ = $-$3 model$^a$}} 
\startdata
Surf. density prof. ($\alpha$)  &  $-2.5, -1.75, -1.1, 0.75,  0.5$ &   $-0.75$ & $-$1   &  $-$3   \\
R$_{in}$     [AU]          &   4, 5.5,  7.5, 11, 15, 18, 25, 30, 40, 50                          &  18 &   10.5           &   15                 \\
R$_{out}$  [AU]           &  20, 40, 80,  150,  300  &   20 & 30    &   300   \\
M$_{dust}$ [M$_{\odot}$] &  (10, 3, 1, 0.3, 0.1, 0.01,0.001)$\times$10$^{-4}$ &  1.0$\times$10$^{-3}$   & 2.0$\times$10$^{-5}$ & 1.8$\times$10$^{-5}$ \\
Flaring index ($\beta$)      & 1.00, 1.07, 1.15 &   1.15  & 1.1    &  1.1 \\
H$_{50}$ (H at 50 AU)   [AU]          & 4, 6, 8 &   4  & 5       &  6  \\
a$_{max}$ [mm] &  0.01, 0.03, 0.1, 0.3, 1, 3, 10, 30 & 3 & 3  & 3  \\
\enddata
\tablecomments{$^a$ The best-fit models fixing the value of $\alpha$ to $-1$ or $-3$ and fine-tuning the other parameters by hand guided by 
the results from the grid.}
\end{deluxetable}

\begin{figure}
\includegraphics[angle=0, width=18.cm, trim=11mm  15mm 1mm 15mm]{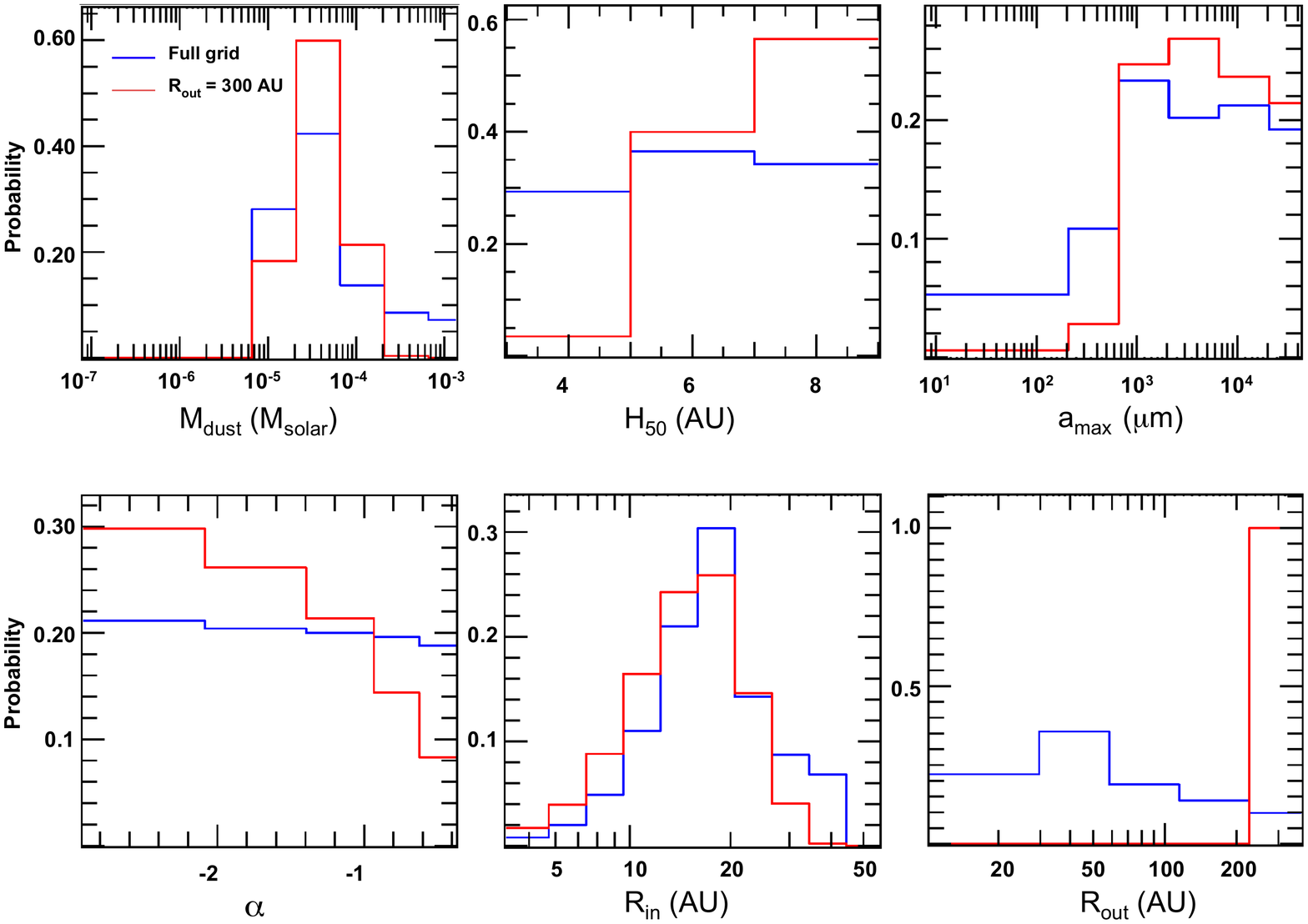}
\includegraphics[angle=0, width=19.cm, trim=11mm 132mm 2mm 13mm, clip]{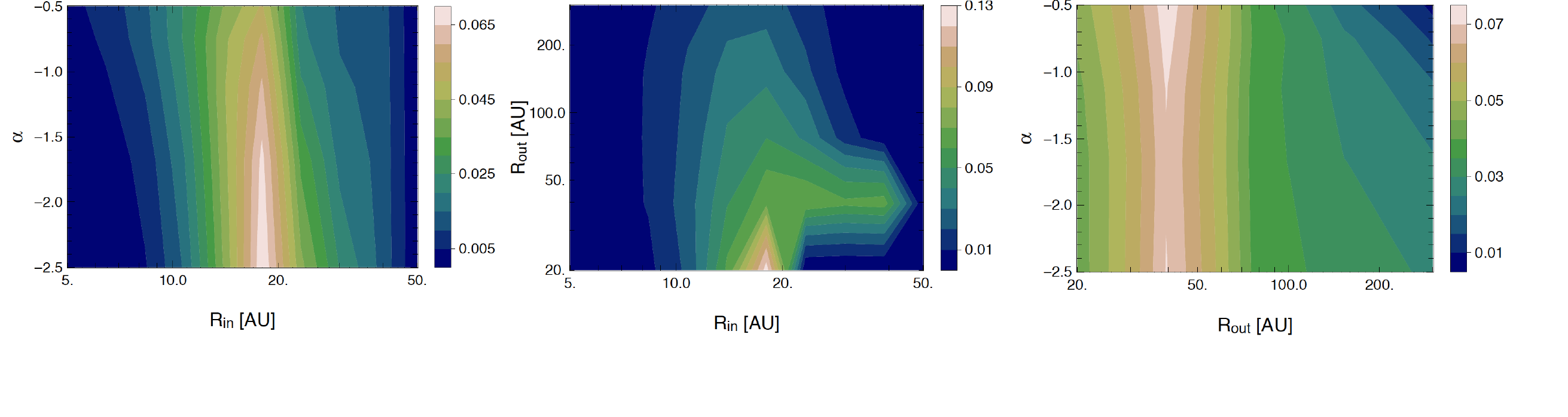}
\caption{ \small \textbf{Top Rows:} Bayesian probabilities for 6 of the 7  free parameters in our model
for the entire grid (blue histograms) and the subset of models with R$_{out}$ = 300 AU (red histograms).
Due to space  limitations, the $\beta$ parameter is not included, but it  presents a rather flat probability distribution.
Models with large R$_{out}$ clearly favor steep surface density profiles and larger H$_{50}$ values. 
\textbf{Bottom Row}: 2--dimensional probability maps of the full grid. 
The highest probability regions are indicated in bright colors. 
Many models with R$_{in}$ $\sim$ 18 AU 
and a steep surface density profile ($\alpha \lesssim -1.5$) fit the data relatively well (left column). Similarly,  
a large number of narrow rings (R$_{in}$ $\sim$ 18 AU and R$_{in}$ $\lesssim$ 50 AU) can reproduce
the observed SED (middle column).  Models with R$_{out}$ $\sim$ 40 AU are most favored by the data.  
Disks with shallow surface density profiles ($\alpha \gtrsim -1.0$) and large outer radius (R$_{out}$ $\sim$ 150-300 AU) 
result in poor fits to the SED (right column). 
}
\label{fig:probs}
\end{figure}

\end{document}